# Interfacial Polarons Driven by Charge Transfer In WSe$_2$/Cuprate Superconductor Systems

Huimin Liu[1,#], Tong Yang [2,#], Xiongfang Liu[1,#], Shengwei Zeng[3], Muhammad Fauzi Sahdan[4], Wenjun Wu[1], Shuo Sun[1], Tengyu Jin[1], Chuanbing Cai[1], Ariando Ariando[4], Mark B. H. Breese[4,5], Wenjing Zhang[6], Andrew T. S. Wee[4,7], Chi Sin Tang[5,*], Ming Yang[2,*], Xinmao Yin[1,*]

[1]Shanghai Key Laboratory of High Temperature Superconductors, Institute for Quantum Science and Technology, Department of Physics, Shanghai University, Shanghai 200444, China

[2] Department of Applied Physics, The Hong Kong Polytechnic University, Kowloon, Hong Kong, 100872, China

[3]Institute of Materials Research and Engineering (IMRE), Agency for Science, Technology and Research (A*STAR), 2 Fusionopolis Way, Innovis #08-03, Singapore 138634, Republic of Singapore

[4]Department of Physics, Faculty of Science, National University of Singapore, Singapore 117542

[5]Singapore Synchrotron Light Source (SSLS), National University of Singapore, Singapore 117603

[6]SZU-NUS Collaborative Innovation Center for Optoelectronic Science & Technology, Key Laboratory of Optoelectronic Devices and Systems of Ministry of Education and Guangdong Province, College of Optoelectronic Engineering, Shenzhen University, Shenzhen 518060, China

[7]Centre for Advanced 2D Materials and Graphene Research, National University of Singapore, Singapore 117546

[#]These authors contributed equally to this work.

*To whom correspondence should be addressed: **Chi Sin Tang**：slscst@nus.edu.sg (C.S.T.);

**Ming Yang**：kevin.m.yang@polyu.edu.hk (M.Y.);**Xinmao Yin**：yinxinmao@shu.edu.cn (X.Y.)

**Author Contributions:** H.L., T.Y., and X.L. contributed equally to this work. X.Y. conceived the project. H.L., X.L. and C.S.T. performed the PL, Raman and SE experiments and analyzed the data. T.Y., and M.Y. performed the DFT calculations and contributed to the theoretical interpretations. H.L., T.Y., X. L, and C.S.T. wrote the manuscript, with input from all the authors.

**Competing Interest Statement:** The authors declare no competing financial interest.

**Abstract**

Understanding the electronic properties of doped copper-oxygen planes remains a significant challenge in condensed matter physics and is crucial to unraveling the mechanisms behind high-temperature superconductivity in cuprates. Recently, the observation of charge transfer and interfacial polarons in superconducting interface has aroused extensive research interest. However, experimental data to investigate charge transfer on the $CuO_2$ plane and the presence of polarons are still missing. Here we conduct extensive research on the optical and electronic properties of two-dimensional material supported on copper-based superconductors. Unlike monolayer-$WSe_2$ on other substrates, monolayer-$WSe_2$ on $La_{1.85}Sr_{0.15}CuO_4$ ($WSe_2$/LSCO) produces a special band structure. Using high-resolution spectroscopic ellipsometry and density functional theory calculation methods, the special electronic structure can be attributed to the formation of the interfacial small polaron at the $WSe_2$/LSCO interface which is driven by charge transfer between the $CuO_2$ plane of the cuprate superconductor and $WSe_2$. In addition, the structural phase transition of the LSCO substrate was observed to reduce the electron-hole (e−h) interaction of $WSe_2$. These findings may spur future investigations on the effect of the interfacial polaron on the superconductivity of cuprates, and highlight the significant influence of interface effects on the electronic structure of $WSe_2$ films. It provides an effective method to further explore the intrinsic relationship between interfacial polarons and superconductivity.

**INTRODUCTION**

The pursuit of high-temperature superconductivity has captivated researchers, with copper oxide based unconventional superconductors at the forefront of intense scientific inquiry [1-3]. Among these materials, the $CuO_2$ planes have received considerable attention due to their crucial role in facilitating superconductivity [4-7]. Despite this focus, debates persist regarding the fundamental physics governing high-temperature cuprate superconductors, particularly in their normal states [8-10]. Central to this debate is the intricate interplay of charges on the $CuO_2$ planes, which has profound implications for the onset and maintenance of superconductivity. While the significance of charge-transfer processes within the $CuO_2$ planes is widely acknowledged, the underlying mechanisms remain a topic of active investigation [11-13].

Polarons are quasiparticles that arise due to the presence of intricate interactions between fermionic particles and bosonic fields [14]. Moreover, two different materials are combined into a heterojunction system, interface polaronic coupling may occur. This can regulate the properties of the heterojunction system [15-17]. For example, the dynamical interfacial polarons formed at the FeSe /$SrTiO_3$ interface are found to play a vital role in enhancing the electron correlation in the overlaying FeSe layer [17, 18]. This enhancement phenomenon has also been found at other interfaces, such as FeSe/STO [19], FeSe/$TiO_2$ [20], Graphene/h-BN [21], etc. The inextricable link between electronic properties of materials and quasiparticle interactions has been further promoted in different material systems [18, 22]. Two-dimensional (2D) materials, particularly metal dichalcogenides, have garnered significant attention owing to their fascinating physical properties and potential applications in next-generation electronics, optoelectronics, and topological quantum devices [23]. On the other hand, interfacing 2D materials with cuprate superconductor surfaces may induce various exotic effects. As the superconducting substrates can interact with the overlaying 2D materials through mechanisms such as strain, charge transfer, electric polarization, and magnetism. So 2D materials is an ideal candidate for exploring interfacial phenomena due to its remarkable properties, including a large direct bandgap, easy exfoliation, tunable electronic structure, and susceptibility to charge transfer [23-25]. These findings provide compelling evidence for the potential existence of interfacial polarons in two-dimensional/cuprate heterostructure systems. This has direct implications for gaining insight into the underlying mechanisms that cause superconductivity and other quantum effects to occur, particularly in the functionalization of emerging quantum systems [26, 27].

Although experimental evidence of polarons in bulk three-dimensional materials is abundant, they have rarely been observed in two-dimensional crystals [28-30]. Small polaron, due to their stronger electron-phonon coupling, can significantly influence and regulate various properties of two-dimensional materials, including the spin and valley degrees of freedom, superconductivity, and energy gaps [31, 32]. We also need to understand how to manipulation of the lattice and electronic structure through an external medium ( such as a substrate) might induce different quantum-mechanical effects at the interface or the two-dimensional layer itself [33].

Here, we report the observation of 2D interfacial small polarons at the heterointerface between monolayer $WSe_2$ and optimally doped $La_{1.85}Sr_{0.15}CuO_4$ (LSCO). There are interfacial lattice strain and charge transfer at the heterointerface, and 2D interfacial small polarons have been found (Figure 1a). We further find that the interfacial effect has a significant impact on the band nesting of $WSe_2$. By utilizing the capabilities of spectroscopic ellipsometry [34], we investigate the electronic properties and quasiparticle dynamics. This approach provides crucial insights into the optical response and the identification of interfacial small polarons at the $WSe_2$/LSCO interface. In particular, detailed analyses indicate that the $CuO_2$ planes within LSCO are responsible for the interfacial charge transfer from monolayer-$WSe_2$ into LSCO. These comprehensive experimental and theoretical investigations suggest that strong charge transfer and interfacial strain are likely the key mechanisms driving the formation of interfacial small polarons. This study indicates the important influence of interfacial effects on the electronic structure of $WSe_2$ thin films. Furthermore, the revelation of interfacial small polarons in this work can stimulate further research to explore the interplay of superconductivity and polaron physics.

**RESULTS AND DISCUSSION**

**Sample Synthesis and Characterization:** Large-area monolayer-WSe$_2$ was synthesized on sapphire (Al$_2$O$_3$) substrate by the chemical vapor deposition (CVD) method using WO$_3$ and Se powders as the reactants [35] and then transferred onto the La$_{1.85}$Sr$_{0.15}$CuO$_4$/LaAlO$_3$ (LSCO/LAO) system using chemical etchant-assisted wet transfer method (details are provided in the Sample Preparation Section). Meanwhile, temperature-dependent resistivity measurements of the ~87.0 nm LSCO/LAO indicates a superconducting transition ($T_C$) at 22 K (Figure S1), consistent with previous reports [1]. Figures. 1b and c compare the Raman and photoluminescence (PL) spectra of monolayer-WSe$_2$ on Al$_2$O$_3$ (WSe$_2$/Al$_2$O$_3$) with that of the monolayer-WSe$_2$ transferred onto the LSCO/LAO film. In the WSe$_2$/Al$_2$O$_3$ system, two main Raman-active modes are observed: (i) The $E_{2g}^1$-mode (~248 cm$^{-1}$) associated with in-plane W-Se bond vibrations, and (ii) the $A_{1g}$-mode (~259 cm$^{-1}$) originating from out-of-plane Se-Se anti-symmetric vibrations. The positions of these signature modes are consistent with previous reports for monolayer-WSe$_2$ [36-38]. Marginal changes in the shape and position shifts of the principal Raman modes can be observed for monolayer-WSe$_2$ on LSCO/LAO. In particular, the broadening of the $E_{2g}^1$-mode could be attributed to its widening differences between the constituent $E_{2g}^{1+}$-mode at a higher wavenumber and the $E_{2g}^{1-}$-mode at the lower wavenumber [39]. Fitting analysis of the original $E_{2g}^1$-mode (249 cm$^{-1}$) for the WSe$_2$/LSCO sample indicates a split into the $E_{2g}^{1-}$ and $E_{2g}^{1+}$ modes, where the former undergoes a redshift to ~242.52 cm$^{-1}$ while the latter undergoes a blueshift to 250.47 cm$^{-1}$ (Figure S2 and Supplementary Information Table. 1). The redshift in $E_{2g}^{1-}$-mode suggests the presence of interfacial tensile strain on monolayer WSe$_2$ at the WSe$_2$/LSCO interface [40]. The out-of-plane $A_{1g}$-mode does not respond to lattice strain but is sensitive to charge doping. Hence, this redshift in the $A_{1g}$ Raman mode is the characteristic of doped MX$_2$(M=Mo/W, X=S/Se) and this have been observed in previous studies [40-42]. In WSe$_2$/LSCO, we observe a ~0.68 cm$^{-1}$ blueshift of the $A_{1g}$ mode to ~259.68 cm$^{-1}$ (Figure S1a), which indicates the possible effect of charge doping due to its sensitivity to a stronger electron–phonon coupling [43-46]. The observed $A_{1g}$-mode redshift provides clear evidence of charge transfer at the WSe$_2$/LSCO interface.

The PL measurement comparison between pristine WSe$_2$/Al$_2$O$_3$ and WSe$_2$/LSCO exhibits a redshift in the characteristic excitonic transitions from the original ~1.67 eV to ~1.57 eV in WSe$_2$/LSCO. This may be attributed to a reduction in band gap due to tensile strain [47, 48]. Besides the redshift, the PL line shape changes significantly. Therefore, we perform deconvolution analysis of the PL spectra peaks into their corresponding radiative recombination components (detailed fitting parameters can be found in the Supplementary Information). The neutral exciton $X^0$ is the ground state of a charge neutral system and trions $X^T$ exciton are the combination of $X$ with one electronic, represented as $e+X^0 \rightarrow X^T$ [49]. It is clear that the presence of excess charge directly controls the intensity of trion emission [50]. Therefore, charge doping will modulate the relative concentration of trions and neutral excitons in monolayer $WSe_2$ [51, 52]. To better quantify our findings, the data is fitted with a multipeak model. As shown in Figure S3, the fitting results indicate a significant decrease in the proportion of $X^T$ on $WSe_2$/LSCO. This indicates a decrease in the electron population of $WSe_2$. By this, we can infer the onset of charge transfer taking place at the interface. The pronounced changes in the Raman and PL spectra of monolayer $WSe_2$ upon its transfer onto the LSCO clearly indicate the P-type doping effects of LSCO on $WSe_2$. It is worth noting that no defect-induced characteristic peaks appear in the Raman spectra of monolayer $WSe_2$ after transfer, and no defect peaks are observed in the PL spectra (Figure.1a and b). The consistent characterization results indicate that the mechanical transfer process did not introduce significant defects. In addition, the subsequent spectroscopic ellipsometry analysis further confirms the high-quality characteristics of the sample. The spectroscopic ellipsometry of $WSe_2$ after transfer is highly consistent with the original monolayer $WSe_2$, and there are no abnormal absorption characteristics within the bandgap caused by defects (such as Urbach tail extension or localized absorption peaks) [53, 54]. This provides a reliable sample foundation for subsequent research on interface properties.

**High-resolution Spectroscopic Ellipsometry Characterization.** High-resolution spectroscopic ellipsometry (SE) was conducted to systematically investigate the changes in the optical and excitonic properties that arise due to the interfacial interaction between $WSe_2$ and LSCO [55]. Figure 2a compares the optical conductivity, $\sigma_1$, of $WSe_2$/LSCO and the as-prepared $WSe_2$/$Al_2O_3$ at 77 K. The shape of the $\sigma_1$ spectrum for the as-prepared monolayer-$WSe_2$ is consistent with results reported in other optical characterization studies [35, 56-58]. Specifically, it exhibits peaks A, B and C located at ~1.67, ~2.12 and ~2.36 eV, respectively. In the high energy region, a broad band nesting feature labelled D is observed at ~2.88 eV. Similar to the PL results, features A and B can be attributed to the excitonic transitions at the K/K' points in the Brillouin zone [53, 59]. The optical bump feature C may be attributed to the resonant exciton that is formed due to the presence of strong electronic correlations in energy bands above the optical band gap [53]. Meanwhile, the broad feature D is ascribed to the optical transitions in the vicinity of the $\Gamma$ and $M$-points of the Brillouin zone where the conduction and valence bands are nested [60].

Significant modifications can be observed for the $\sigma_1$ spectrum belonging to WSe$_2$/LSCO despite the persistence of the aforementioned optical features which confirms the quality of the monolayer-WSe$_2$ after its transfer onto LSCO (Figure 2a). Variations to the $\sigma_1$ spectrum suggest that the optical properties and electronic structures of monolayer-WSe$_2$ have changed due to interfacial interaction taking place at the WSe$_2$/LSCO interface. Firstly, the intensity of the principle features previously described have undergone an attenuation. Specifically, the band nesting feature D in the photon energy region of ~2.7–3 eV shows a considerable intensity reduction due to changes in the WSe$_2$ band structure induced by interfacial hybridization and lattice strain which in turn reduces its band nesting structure [60]. The interfacial effect leads to the reduction in the band nesting structure (around the $\Gamma$ point) of monolayer-WSe$_2$. Black curves in the band dispersion indicate a good band nesting area (Figure1a). Likewise, each optical feature is redshifted and broadened by different extent (Figure 2a). These changes in spectral features are in agreement with the results elucidated from both the Raman and PL and align with previously reported theoretical predictions, which suggest that the reduction in bandgap due to strain originates from the diminished bond angle of Se−W−Se and subsequent changes in orbital overlap [47, 48].

Previous studies indicate that charge doping could cause a redshift in exciton peaks A and B of monolayer-WSe$_2$ [61], which in turn implies a possible charge transfer at the WSe$_2$/LSCO interface. First-principles calculations were then conducted to scrutinize any onset of interfacial charge transfer process. The Fermi level of CuO-terminated LSCO surfaces is ~1 eV below the valence band maximum (VBM) of monolayer-WSe$_2$ (Figure 2c). This would lead to an interfacial charge transfer from the WSe$_2$ layer to the CuO-terminated LSCO. Hence, this computational analysis substantiates the experimental results which point to charge transfer at the WSe$_2$/LSCO interface (see calculation methods in the Supplementary Information for more details). Peak C is a transition from the valence band maximum to the energy band at K-point. Charge transfer and strain effects cause changes to the energy band at this position, resulting in the same red shift and intensity reduction effect as the A and B exciton peaks. Notably, significant changes were observed in a previously unidentified prominent optical shoulder labeled A* at ~1.46 eV exciton A, and its origins will be discussed thereafter.

**Temperature-dependent High-resolution Spectroscopic Ellipsometry.** Figure 2b displays the optical conductivity, $\sigma_1$, of WSe$_2$/LSCO at temperatures between 77 and 270 K. Apart from a monotonic reduction in peak intensity, the position and width of the principal excitonic and optical features show a red shifting and broadening trend (Figure S4). These temperature-dependent changes are consistent with previous spectroscopic ellipsometry and reflectance experimental studies of both CVD-grown and exfoliated monolayer TMDs [53, 62, 63]. The red shifting and broadening behavior of the excitons with increasing temperature can be attributed to processes related to thermal expansion, strengthened electron-phonon interaction, and radiative recombination in monolayer-WSe$_2$ [64, 65]. In particular, the nonradiative electron-hole recombination process is enhanced as temperature increases and this reduces the probability of radiative transition in monolayer-WSe$_2$. This contributes significantly to the temperature-dependent variations in the intensity and peak widths that are observed.

Having analyzed the $\sigma_1$ spectra, we proceed to discuss the temperature-dependent changes in spectral weights (SW) [34]. Firstly, emphasis is on the low-energy region (1.30 – 2.18 eV) where the key optical structures are present. To further analyze the temperature-dependence of each optical feature, the photon energy range is further sub-divided into three principal spectral regions where the respective excitons and the additional optical feature A* have been observed. Namely, regions I (1.30–1.48 eV), II (1.48–1.89 eV) and III (1.89–2.18 eV), respectively (See Supplementary Information for details). The SW of the respective optical region can be calculated based on the following expression,

$$SW = (\pi N_{eff} e^2) / (2m_e) = \int_{E_1}^{E_2} \sigma_1(E) \, dE$$

where $N_{eff}$ denotes the electron density, $e$ denotes the electronic charge, and $m_e$ refers to the electron mass. As shown in Figure 2d, the SW of monolayer-WSe$_2$/Al$_2$O$_3$ in low-energy regions and total region (1.30~3.70 eV) increases linearly with temperature. The changes of the number of the effective carriers are consistent with the spectral weight behavior (Figure.S7). This means that the total effective electron density of WSe$_2$ increases over the entire spectral range, and the number of charges involved in the optical transition increases with temperature [66, 67].

However, the SW of monolayer WSe$_2$/LSCO in total energy region and low-energy regions (I, II, III)

increases first and then decreases around 140 K [68]. It is worth noting that SW of WSe$_2$/LSCO is similar to the structural transition temperature of LSCO (Figure S5). This indicates that the $N_{eff}$ mutation in the region containing A*, A and B excitons at ~140 K may be related to the structural phase transition of LSCO. Conversely, the spectral weight in the high energy region where the band nesting feature is present (~2.18-3.70 eV) shows a monotonic increase with temperature (See Supplementary Information for details). The temperature-dependent behavior of the spectral weight (or $N_{eff}$) in the high energy and the low energy region shows opposite behaviors. This could possibly be attributed to the structural phase transition of LSCO has a significant influence on the direct band gap of WSe$_2$ at the *K*-point resulting from a change in interfacial lattice strain. The structural transformation has no significant effect on the band nesting structure in the high-energy region [60].

The Feature A* observed in WSe$_2$/LSCO heterostructures persists across the entire temperature range (77–270 K). Repeated optical measurements conclusively demonstrate that this feature originates neither from optical artifacts nor defect states in monolayer WSe$_2$. Its presence throughout the entire temperature range further confirms that structural transformation of LSCO is not the cause of this feature. In addition, monolayer WSe$_2$ on LSCO is under smaller tensile stress as compared with that when it is on the Al$_2$O$_3$ substrate. We can thus eliminate the interfacial lattice strain as the contributing factor to the appearance of feature A* in optical region I. While interlayer excitons appear at a lower energy position compared to excitons from the actual material [69-71], the LSCO layer is in metallic state and does not satisfy the conditions for type-II band alignment with WSe$_2$ which is a necessary condition for the formation of interlayer excitons [70, 72]. Besides, there is no appearance of excitonic signals in the low-energy region of the PL peak (Figure 1c) which further rules out this possibility. In what follows, we will show that it can be attributed to the interfacial small polarons at the interface between the LSCO and WSe$_2$ layers.

**Bryksin Small-Polaron Model**. To further investigate the physical properties of optical feature A* located below exciton A observed in WSe$_2$/LSCO in the photon energy region between 1.30 and 1.48 eV, curve fitting is performed using the Bryksin small-polaron model (See Supplementary Information for details) which models the electron hopping between in-plane neighboring sites upon activation (Eq. 1 in the Supplementary Information) and its temperature-dependent behavior is analyzed [73, 74]. The optical responses of the polaron are compatible with the theoretical small-polaron model as seen in the stacked temperature-dependent $\sigma_1$ spectra in Figure 3a. As shown in Figure S8, the residual plot results are consistent with the normal distribution, verifying that the assumptions of the Bryksin model are reasonable. The principal parameters derived from this analysis include the polaron hopping energy, $E_a$, polaron bandwidth, $\Gamma$, and phonon energy, $E_{LO}$, as displayed in Figures. 3b-d, respectively.

Between 77 and 270 K, the polaron hopping energy between adjacent ion sites (e.g., $W^{2+} \rightarrow W^{2+}$) in WSe$_2$/LSCO interface exhibits a monotonically increasing trend with increasing temperature from ~0.368 eV at 77 K to ~0.381 eV at 270 K (Figure 3b). This increase may be attributed to the spreading polaron wave function as temperature increases, which allows for a greater ease of interstitial charge hopping [75]. Meanwhile, the polaron bandwidth, $\Gamma$, also displays a progressive broadening trend with temperature from ~0.0687 eV at 77 K to ~0.092 eV at 270 K. This is consistent with the behavior of small polarons which corresponds to the phonon broadening trend of the local electronic energy levels and this in turn progressively leads to the broadening of the absorption bands [75]. Specifically, in WSe$_2$/LSCO, the small-polaron absorption band broadens with increasing temperature by thermal effects [76]. The phonon energy, $E_{LO}$, presents a similar temperature-dependent trend with $E_a$. As the temperature increases, the lattice vibrations become more intense with rising thermal energy. Nevertheless, the phonon energy remains limited below ~15 meV, positioning it relatively close to the acoustic phonon mode of monolayer-WSe$_2$ and the LSCO substrate [68, 77]. This interfacial polaron formation may be attributed to electron-phonon interactions involving transferred holes and the overlapping acoustic phonon modes at the WSe$_2$/LSCO interface.

Although defect states or interfacial charge transfer may contribute to low-energy optical response, PL characterization (Figure 1c) and spectroscopic ellipsometry experiments (Figure 2b) can effectively rule out their dominant role. There is no new band generated at the peak position. The temperature sensitivity of feature A * is quantitatively consistent with the polaron of the Bryksin model, further supporting its electron-phonon coupling origin. Overall, the compatibility between the optical features observed in the respective $\sigma_1$ spectra and the Bryksin small-polaron model strongly suggests that these optical responses genuinely reflect the presence of small polarons at the WSe$_2$/LSCO heterojunction.

**DFT Results of the WSe$_2$/LSCO Interface**. DFT calculations were further performed to shed light on the origins of the small polarons at the WSe$_2$/LSCO interface. There are two potential origins of the observed polaron and that they might be hosted (i) solely by the *p*-doped WSe$_2$ on LSCO, or (ii) at the

WSe$_2$/LSCO interface. In the latter case, excess holes in monolayer-WSe$_2$ or near the interface could be dressed by the phonons of the LSCO surface according to a penetrating-field mechanism to form interfacial polarons, as observed in other systems such as the SnSe$_2$/SrTiO$_3$, FeSe/SrTiO$_3$ and MoS$_2$/TiO$_2$ interfaces[15-17]. Based on our DFT calculations as elaborated below, the first proposed origin can be ruled out and this small polaron is more likely the interfacial polaron.

In addition to the charge transfer which leads to the *p*-doping of monolayer-WSe$_2$, the supported WSe$_2$ is subjected to structural strain due to lattice mismatch with the LSCO layer. Figure 4a displays the projected band structure of the strain-free WSe$_2$, where the valence band maximum (VBM) is located at the *K*-point. The Bloch states around VBM at the *K*-point are mainly from the laterally-oriented W-$d_{xy}, d_{x^2-y^2}$ orbitals, while those at the -point are mainly from the W-$d_{z^2}$ orbital. Therefore, the application of tensile strains would shift the energy band at the *K*-point downwards and the VBM will be shifted to the -point under a tensile strain of at least 4% (Figure 4b). Along with the change in VBM position, the hole effective mass is also greatly enhanced, thereby implying that the excess holes could be self-trapped and localized with greater ease in the WSe$_2$ layer under tensile strain. By applying a recently developed theoretical model, it can be estimated whether the Fröhlich interaction can stabilize the hole polaron in WSe$_2$ [33]. The inequality has been developed as a critical condition for the formation of polarons in 2D materials:

$$\beta = \frac{\epsilon_{ion} h_*^m d}{m_0 a_0} > 2$$

where $\epsilon_{ion}$, $m_h^*$, $d$, $m_0$ and $a_0$ denote the ionic contribution to the dielectric constant, the hole effective mass, the effective thickness, the bare electron mass and the Bohr radius, respectively.

As displayed in Figure 4c, the inequality is satisfied for monolayer-WSe$_2$ when the tensile strain is at least 4%. This is indicative of the existence of the hole polarons in WSe$_2$ under tensile strains. However, with the estimated polaron radius above ~300 Å, the quantum coherence between the electrons and phonons is unlikely to be maintained over such a distance and neither will it be experimentally observable. Attempts were also made to explicitly simulate the hole polaron in both strain-free and 4%-strained WSe$_2$ supercell at the HSE06 level. However, the excess hole is unable to induce a significant local lattice distortion and be self-trapped in the WSe$_2$ layer (cyan iso-surface in Figure 4d). Based on these analyses,

WSe$_2$ alone is unlikely to host a stable hole polaron. Therefore, the experimentally observed polaron could be rooted at the WSe$_2$/LSCO interface. Wherein, the additional coupling between the holes within WSe$_2$ or near the interface and the phonons of the LSCO substrate may play a vital role in further stabilizing/localizing the hole polarons.

The SE spectra of monolayer WSe$_2$/LSCO show a peak A* at 1.5eV which can be well interpreted by interfacial small polarons developed at the interface between WSe$_2$ and LSCO. Recent studies in FeSe/STO, LaAlO$_3$/SrTiO$_3$ have found that interfacial polarons enhance the superconductivity at the interface[17, 18]. The appearance of polarons at this WSe$_2$/LSCO heterojunction interface encourages us to compare this interface with similar interfacial superconducting systems. And solid evidence has been recently discovered that interfacial polarons play critical roles in the interfacial conductivity and even enhanced interfacial superconductivity.

Conventional theory holds that polaron formation originates from long-range electron-phonon interactions, which are perceived to facilitate superconducting pairing. However, the precise role of small polarons in the pairing mechanism remains a scientific question to be addressed. Recent experimental and theoretical studies [78-83] have revived interests in the phonon contribution to the pairing mechanism in cuprate superconductors, particularly in the context of strongly correlated electron systems. A theoretical study has suggested that small polarons may promote localized electron pair formation through local electron-phonon coupling [84], thereby offering an alternative pathway for superconducting pairing.

This work reports the direct observation of interfacial polarons coupled to CuO$_2$ planes, providing experimental insights into phonon-mediated pairing mechanisms in strongly correlated systems. Notably, the interface-engineered small polarons may modulate local electron-phonon interactions, potentially influencing either the pairing strength or coherence. Understanding the interplay between superconductivity and polaron physics remains highly challenging, and we anticipate that revealing interfacial small polarons in this work may encourage further investigations in this direction.

**CONCLUSIONS**

In conclusion, we prepared a large monolayer of WSe$_2$ and transferred it onto a copper-oxide high-temperature superconductor to investigate the properties of the WSe$_2$/LSCO interface. Our observation indicates that the WSe$_2$ on LSCO substrates exhibit a distinctive interface effect compared to on Al$_2$O$_3$ substrates. Through a comprehensive investigation that comprises temperature-dependent spectroscopy ellipsometry, Raman spectra, photoluminescence spectra, and DFT calculations, we determined that charge transfer between the CuO$_2$ plane of LSCO and monolayer-WSe$_2$ drives the formation of 2D interfacial small polarons. Our work emphasizes the critical role of charge transfer, strain, and interfacial electron-phonon coupling in mediating the formation of these polarons within the WSe$_2$/LSCO heterostructure. This study is pivotal in advancing our understanding the electronic structures of the CuO$_2$ plane and for exploring the mechanisms underlying high-temperature superconductivity. It provides important insights into the fundamental physical principles of superconductivity and complex interfacial phenomena. The discovery of interfacial small polarons could stimulate further research to explore the interaction of superconductivity and polaron physics. This study is pivotal in advancing our understanding of the electronic structures of the CuO$_2$ plane and in exploring the mechanisms underlying high-temperature superconductivity.

**Materials and Methods**

**LSCO:** High-quality thin-film La$_{1.85}$Sr$_{0.15}$CuO$_4$ samples were grown on atomically smooth [001]-oriented LaAlO$_3$ single-crystal substrates, using pulsed laser deposition. The pure cation oxides powders of La$_2$O$_3$ (99.999%), Sr$_2$O$_3$ (99.997%), and CuO (99.9999%) are used for the preparation of the ceramic target. According to the chemical formula, the materials were weighed and mixed according to the appropriate chemical stoichiometry. Note that the 87.0nm-LSCO/LAO sample is superconducting with T$_c$~22 K. (Figure S1). This is in good agreement with a previous substrate and thickness-dependent study on the T$_c$ of optimal-doped LSCO films brought about by the effects of epitaxial strain [85].

**WSe$_2$:** Atomic layers of WSe$_2$ were synthesized on a sapphire (Al$_2$O$_3$) substrate employing the chemical vapor deposition (CVD) technique, utilizing WO$_3$ and Se powders as reactants. The process involved an annealing step at 200°C to promote intimate contact between the film and substrate while removing any residual impurities. Raman and photoluminescence (PL) spectra WSe$_2$ monolayers are provided in Supplementary Information Figs.S2 and S3. Comparative analysis with recent reports on monolayer WSe$_2$ and monolayer WSe$_2$ on sapphire reveals striking spectral similarities, affirming the high quality of the samples as monolayers.

**Chemical etchant-assisted wet transfer methods:** The transfer of monolayer-WSe$_2$ onto the LSCO layer is conducted using polymethyl methacrylate (PMMA). In which case, the PMMA film is spun onto the surface of a WSe$_2$/Al$_2$O$_3$ stack to serve as a mechanical support layer and is left overnight to ensure strong adhesion between the PMMA and the 2D film. Subsequently, the entire stack is immersed in hot deionized (DI) water at an angle of approximately 45° to the water surface. After a few seconds, water begins to penetrate the film/substrate interface due to capillary forces, leading to the delamination of the PMMA/2D film from the substrate[86]. The delaminated film then floats to the surface and is transferred onto an 87.0 nm LSCO/LAO substrate. Following this, the PMMA/WSe$_2$/LSCO is cleaned using DI water baths and transferred onto the target substrate, after which an acetone bath is employed to dissolve the PMMA support layer.

**Data Availability Statement:** The data that support the findings of this study are available from the corresponding author upon reasonable request.

**Supporting Information:** Detailed methodology and analysis of the spectroscopic ellipsometry data, Bryksin small-polaron model and transport measurement data, and additional DFT calculations. Supporting Information is available from the ACS Nano website at DOI: 10.1021/acsnano.5c05277

**Acknowledgements**

This work was supported by National Key R&D Program of China (Grant No. 2022YFE03150200), the National Natural Science Foundation of China (Grant Nos.52172271, 12374378, 52307026), Shanghai Science and Technology Innovation Program (Grant No. 22511100200, 23511101600). M. Y. acknowledges the funding support from Hong Kong Research Grants Council (project no.: P0046939 and P0045061) and The Hong Kong Polytechnic University (project no.: P0050570, P0048122 and P0047956). C.S.T acknowledges the support from the NUS Emerging Scientist Fellowship. The authors acknowledge the computing resource from the National Supercomputing Centre, Singapore (https://www.nscc.sg).This research is supported by the Ministry of Education, Singapore, under MOE-T2EP50124-0003.

**Figures and Tables**

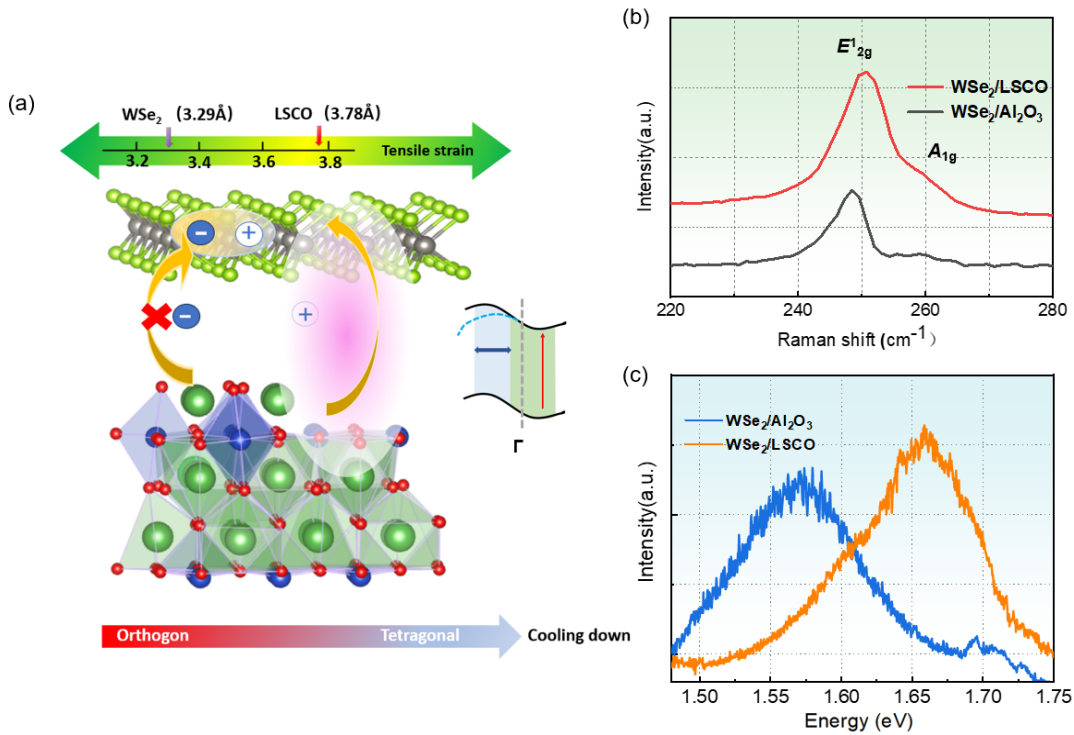

Figure1. (a) Crystal model diagrams of monolayer-WSe$_2$/Al$_2$O$_3$ on 87.0 nm LSCO heterostructure. WSe$_2$ is subjected to tensile strain by the LSCO and a charge transfer occurs with the CuO$_2$ plane, resulting in the formation of interfacial polaron at the interface. This interfacial effect leads to the reduction in the band nesting structure (near the Γ point) of monolayer-WSe$_2$. Black curves in the band dispersion indicate a region where band nesting takes place. The red arrow indicate the direction of the optical transitions, which appear as peak C in the measured optical spectra. The blue arrow indicates the likely band region which is modified, thereby reducing the band nesting structure. (b) The Raman, and (c) photoluminescence spectra of monolayer-WSe$_2$/Al$_2$O$_3$ and WSe$_2$/LSCO.

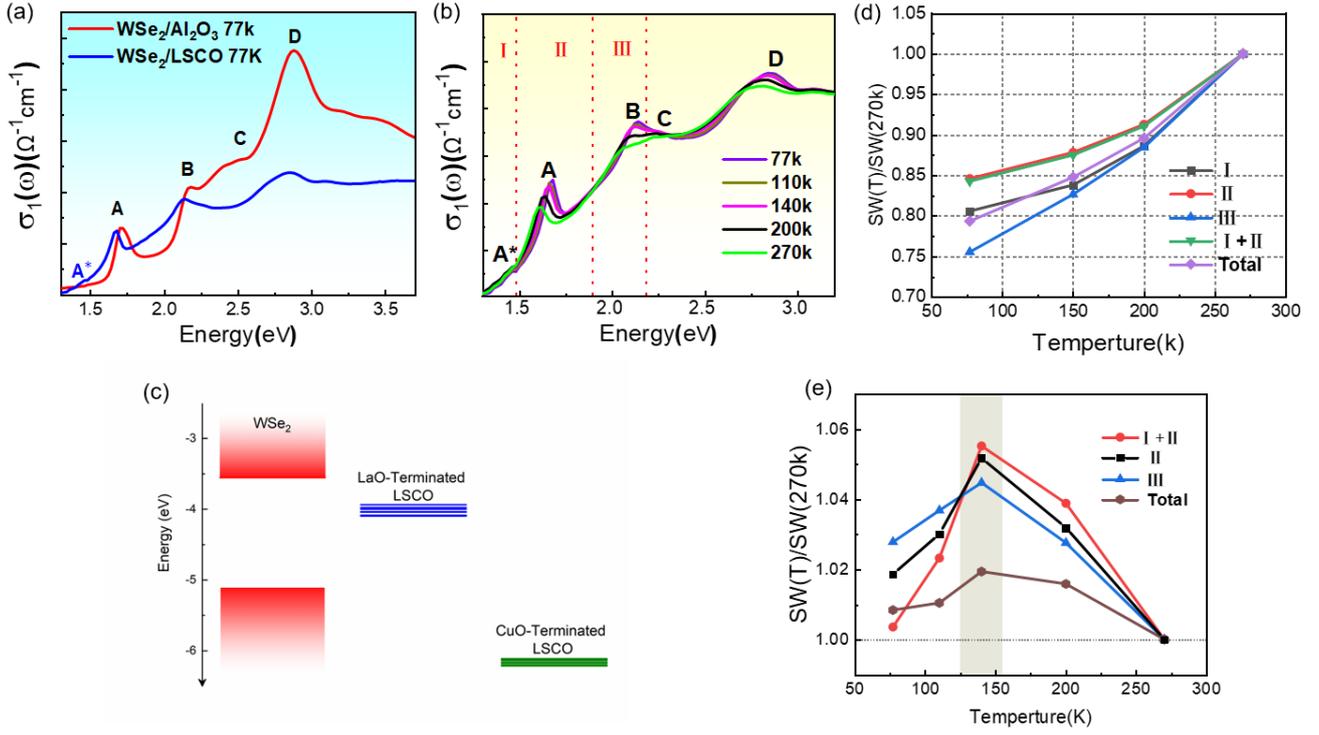

Figure2. (a) Optical conductivity, $\sigma_1(\omega)$ of monolayer-WSe$_2$ and WSe$_2$/LSCO at 77K. (b) $\sigma_1(\omega)$ from 1.3 to 3.2 eV as a function of temperature for WSe$_2$/LSCO. (c) The alignment of the band edges of monolayer WSe$_2$ with the Fermi level of the LaO- and CuO-terminated LSCO surfaces based on Anderson's rule. (d) and (e) Spectral weights of monolayer-WSe$_2$/Al$_2$O$_3$ and monolayer-WSe$_2$/LSCO, respectively. Integrated spectral weight $\frac{SW(T)}{SW(77K)}$ defined as $\frac{\int_{E_1}^{E_2} \sigma_1(E,T)dE}{\int_{E_1}^{E_2} \sigma_1(E,T=77\ K)dE}$ in different spectral regions, I (1.30 – 1.48 eV), II (1.48 – 1.89 eV), and III (1.89 – 2.18 eV).

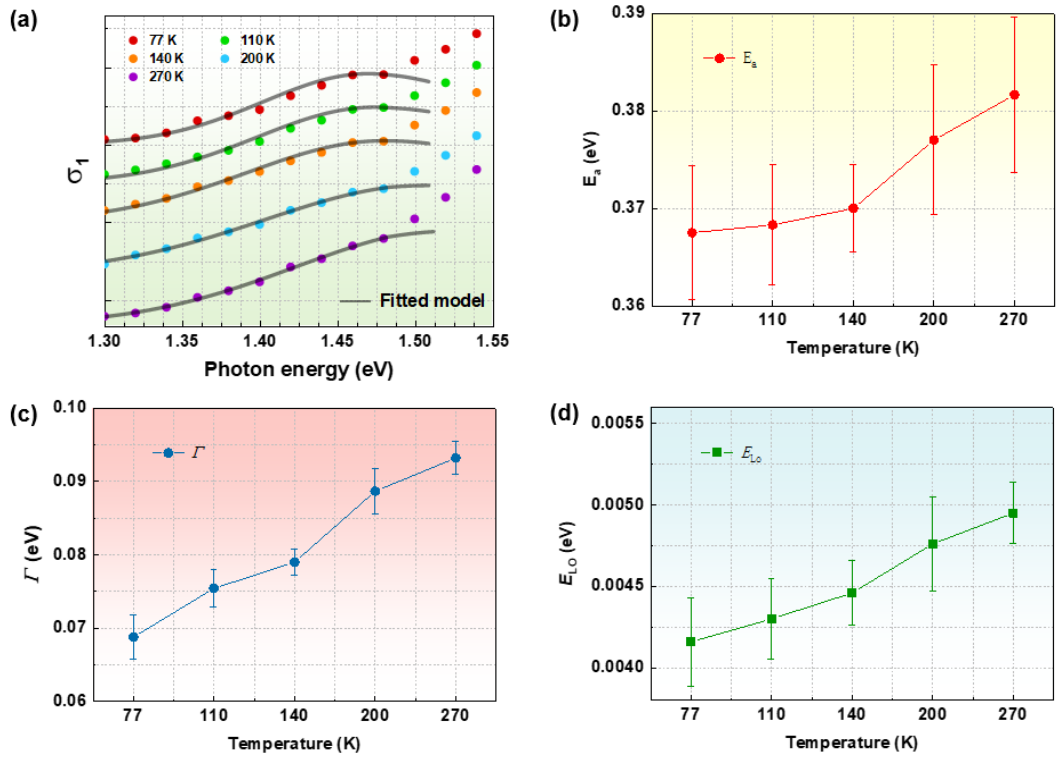

Figure 3. The zoom-in optical conductivity spectra $\sigma_1$ of WSe$_2$/LSCO heterostructure where optical feature $A^*$ is present at different temperatures. (a) Overlay of the optical data and the fitted curve (dashed lines) based on the Bryksin small-polaron model (see supplementary Information). Parameters derived from Bryksin's small-polaron fitting analysis of the WSe$_2$/LSCO heterostructure. (b) Polaron hopping energy, $E_a$, (c) polaron bandwidth, $\Gamma$, and (d) phonon energy, $E_{LO}$.

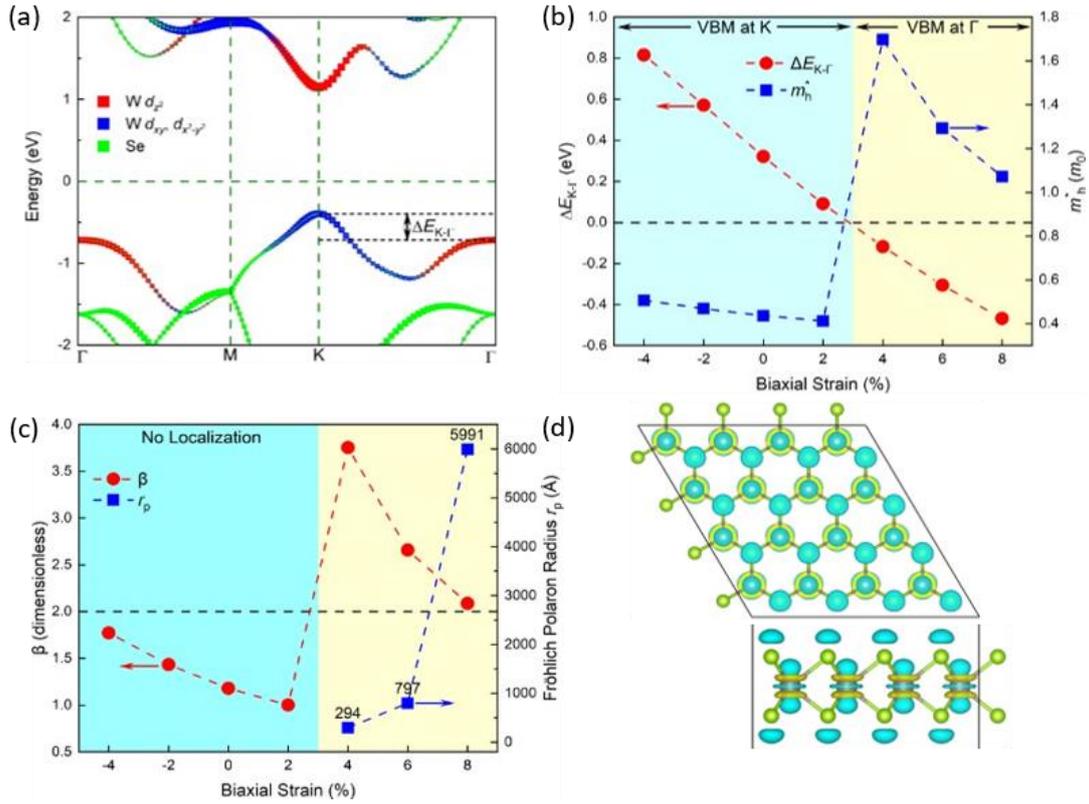

Figure 4. (a) The projected band structure of the strain-free WSe$_2$ at the PBE level of theory. (b) The energy difference between the highest valence state at the $K$- and -points ($\Delta E_{K\text{-}\Gamma}$) and the hole effective mass around the valence band maximum as functions of the applied biaxial strain. (c) The polaron formation descriptor $\beta=\epsilon_{ion}m_h^*/m_0 a_0$ and the Fröhlich polaron radius as functions of the biaxial strain. Polarons do not form when $\beta\leq 2$. (d) The charge density difference between one hole-doped and charge-neutral states of the optimized 4 %-strained 4×4×1 WSe$_2$ supercell of which a W atom was selected and displaced from its equilibrium position before structural relaxation. Yellow and cyan iso-surfaces denote charge accumulation and depletion. The iso-value was set to 0.0005 $e/$Å$^3$.